\documentclass[trackchanges,twocolumn]{aastex7}

\usepackage{xcolor}
\newcommand{\tb}[1]{\textcolor{blue}{#1}}

\renewcommand{\tb}[1]{{#1}}

\begin{document}

\title{Distribution of energy release events due to magnetic braiding}

\author[0000-0002-1089-9270]{David I. Pontin}
\affiliation{The University of Newcastle,
University Dr, Callaghan,
NSW 2308, Australia}
\email[show]{david.pontin@newcastle.edu.au}

\author[0000-0001-8882-1708]{Klaus Galsgaard}
\affiliation{School of Mathematics and Statistics, University of St Andrews, North Haugh, St Andrews, KY16 9SS, Scotland, UK}
\email[show]{klaus.galsgaard@gmail.com}  

\author[0000-0003-2255-0305]{James A. Klimchuk}
\affiliation{Heliophysics Science Division, NASA Goddard Space Flight Center, 8800 Greenbelt Rd., Greenbelt, MD 20771, USA}
\email[]{james.a.klimchuk@nasa.gov}

\begin{abstract}
Energy conversion by reconnection-powered nanoflare heating is one of the leading explanations for the heating of the solar chromosphere and corona.
The aim of this paper is to shed light on this mechanism by exploring the magnetic Reynolds number dependence of the energy conversion process. 
To do this we employ  boundary-driven, magnetohydrodynamic, flux-braiding  simulations at different magnetic Reynolds numbers ($R_m$), and explore in detail the properties of the individual magnetic energy release events. 
The properties of the reconnecting current sheets that mediate the energy release are shown to depend on $R_m$. 
For increasing $R_m$, the current sheets become thinner, more intense, and more numerous.
For sufficiently large $R_m$, the current sheets fragment along their length, leading to a sharp cutoff in the current sheet length distribution. The cutoff is consistent with the threshold for non-linear tearing/plasmoid instability.
For increasing $R_m$ the magnetic field lines become increasingly tangled, the mean and peak values of the magnetic field strength increase, and the Poynting flux into the domain increases, implying that the heating rate also increases.
The global reconnection rate is essentially independent of $R_m$.
These results support the braiding mechanism as a viable way to effectively heat the internal portions of coherent flux tubes in the corona. 
\end{abstract}

\keywords{\uat{Galaxies}{573} --- \uat{Cosmology}{343} --- \uat{High Energy astrophysics}{739} --- \uat{Interstellar medium}{847} --- \uat{Stellar astronomy}{1583} --- \uat{Solar physics}{1476}}


\section{Introduction}\label{sec:intro}

Explaining the heating of the Sun's corona to multi-million degree temperatures remains one of the biggest outstanding problems in solar physics. Numerous potential mechanisms have been proposed for providing the required thermal energy input, and it is likely that different mechanisms dominate in different parts of the solar atmosphere. Here we consider the braiding mechanism, first proposed by Parker \cite[see, e.g.,][]{parker1972,parker1988}. 

In the braiding picture, the convective photospheric flows cause a random walk of the footpoints of magnetic field lines that arch up into the atmosphere. Because these motions are slow compared to the typical Alfv\'en speed in the atmosphere, it is assumed that the magnetic field stays close to an equilibrium, but the field lines become increasingly tangled. At some point, electric current sheets form, and reconnection at these current sheets converts stored magnetic energy both directly to thermal energy, and (first) to kinetic energy. Much early work analysing this mechanism focussed on \cite{parker1972}'s assertion that the current sheets are formed due to the non-existence of a (smooth) equilibrium \citep[see, e.g., the review by][]{pontin2020a}. While there is now a weight of evidence against this assertion, it is well established that the continual action of the photospheric flows produces current layers that grow in intensity and decay in thickness exponentially with time, such that eventual reconnection onset is inevitable \citep{vanballegooijen1988a,vanballegooijen1988b,mikic1989,pontin2015a}.

Many authors have simulated the ``flux braiding'' mechanism, employing a range of approaches. Differences between the simulations include different governing equations (e.g., full MHD, reduced MHD (RMHD)), box dimensions, driving flow spatial and temporal dependence, resistivity and viscosity models. We do not attempt to summarise these here, but instead refer the reader to the review by \cite{pontin2020a}. In almost all cases, loop curvature is ignored, and flux braiding is simulated in a Cartesian geometry with the magnetic field connecting two line-tied, perfectly conducting plates. In the following we refer to this direction of the initial/axial magnetic field as the $z$-direction. 

Some key, general results common to these flux braiding simulations are as follows.
Thin ribbons of current form, whose thickness decreases exponentially in time. The long-time behaviour reaches a statistically steady state where the injection of magnetic energy (and field line tangling) is balanced by dissipation of magnetic energy (and reconnection-mediated untangling). The properties of the statistically steady state depend on factors such as the loop length, driving speed and magnetic field strength \citep[e.g.,][]{dmitruk2003,rappazzo2008}.
The free magnetic energy (i.e., energy of the magnetic field with the background $B_z$ neglected) dominates the kinetic energy, and scales, for example, with the driving speed.
The detailed geometry of the driving flow is argued to have  minimal importance \citep[e.g.,][]{rappazzo2010} but overall driver properties may affect the characteristics of the energy release, such as the helicity injection rate \citep{ritchie2016} or the correlation time \citep{dmitruk1999}.

One critical difference between all of the flux braiding simulations and the solar corona is the magnetic Reynolds number, $R_m$. 
Most studies that consider the scaling of the system with $R_m$ use the RMHD approximation. However, $B_\perp=|{\bf B}_{xy}|$ is expected to grow with $R_m$, meaning that for  sufficiently large $R_m$ the RMHD assumptions may become invalid, unless the growth of $B_\perp$ with $R_m$ saturates, as proposed by \cite{ng2012}.
Studies addressing the $R_m$ scaling typically focus on the net rate of energy dissipation, and argue that this should become independent of $R_m$ \citep[e.g.,][]{galsgaard1996,rappazzo2008}.
One of the main goals of this study is to focus on the $R_m$ dependence of the energy release events in detail.

One of the reasons that assessing the $R_m$ dependence is so critical as that in most models the state of the ``coronal'' volume depends on $R_m$, and the magnetic energy (Poynting) flux into the volume depends on this coronal state.
In order to obtain the correct energy balance to maintain the corona at its observed temperature, \cite{parker1983a} showed that, at least on average, the field should be inclined at an angle of 10--20$^\circ$ to the vertical. It has thus been proposed that the energy release process in the current sheets may ``switch on'' when the magnetic shear across the current layer exceeds a critical angle, due to the onset of a rapidly growing instability.
This idea was developed in a series of papers by \cite{dahlburg2005,dahlburg2009} who simulated a sheared, line-tied magnetic field with a single, planar current layer. In the linear phase of the current sheet instability a series of co-aligned, twisted flux tubes is formed in the current layer. These undergo a ``secondary instability'' -- essentially an ideal kinking of those twisted flux tubes -- for sufficiently large shear across the current layer, leading subsequently to a turbulent evolution. {It is worth noting that the current grows much more rapidly when the single, coherent shear motion is replaced by a sequence of shear flows in different directions, more representative of the complex photospheric flows \citep{mikic1989,galsgaard1996}.}

More recently, \cite{leake2020} have undertaken a series of simulations in a triply-periodic domain for different current sheet lengths and magnetic shear angles. They
identified two distinct regimes of the current sheets, based on the relationship between the sheet length $L$ and the wavelength $\lambda$ of the fastest-growing parallel mode. When $L>\lambda$ sub-harmonics of the fastest-growing mode emerge, with nonlinear interactions and 3D plasmoid coalescence dominating the dynamics.
When $L<\lambda$ no sub-harmonics form, and significant energy conversion occurs only under strong magnetic shear, where oblique modes grow enough to interact with the dominant parallel mode. \cite{leake2024} extended those results by taking the important step of considering current sheets that are dynamically thinning. These ideas provide important context for the fragmenting current layers that we find in our simulations, described below.

The aim of this paper is to explore the Reynolds-number dependence of energy conversion in  boundary-driven flux-braiding simulations, by exploring in detail the properties of the individual magnetic energy release events. The setup of our simulations is described in Section \ref{sec:setup}. In Sections \ref{sec:globalresults} and \ref{sec:sheet_results} we present and discuss our results, before finishing in Section \ref{sec:conc} with conclusions.

\section{Description of simulations}\label{sec:setup}

In line with the vast majority of flux braiding simulations \citep[reviewed in][]{pontin2020a} we ignore loop curvature and solve the MHD equations in a Cartesian domain. 
The code employed is the STAGGER code \citep{galsgaard1997}, which solves the standard induction, momentum, continuity and energy equations for single-fluid MHD in dimensionless form. Gravity, thermal conduction, and radiative losses are all neglected.

Dissipation is handled with a ``hyper''-resistivity and -viscosity, using the formulation described in Equation (9) of \cite{gudiksen2011}, where for the momentum dissipation (viscosity) we have $\{\nu_1,\nu_2,\nu_3\}=\{0.005, 0.01, 0.5\}$ while for the magnetic field dissipation (resistivity) we have $\{\eta_1,\eta_2,\eta_3\}=\{0.005, 0.01, 0.0\}$. These numbers are in general chosen to allow structures to collapse close to the grid scale before starting to diffuse, while also minimising spurious numerical oscillations, and are problem-specific. \tb{Note that the code works in dimensionless units with $\mu_0=1$, so that the magnetic resistivity and diffusivity are equivalent.}

We begin with an initially homogeneous plasma and magnetic field, with (dimensionless) values for the magnetic field components $B_x=B_y=0$, $B_z=1$, plasma velocity ${\bf v}={\bf 0}$, density $\rho =1$ and pressure $p=0.025$ (so that $\beta=0.05$). 
\begin{figure}
    \centering
    (a) \includegraphics[width=0.92\linewidth]{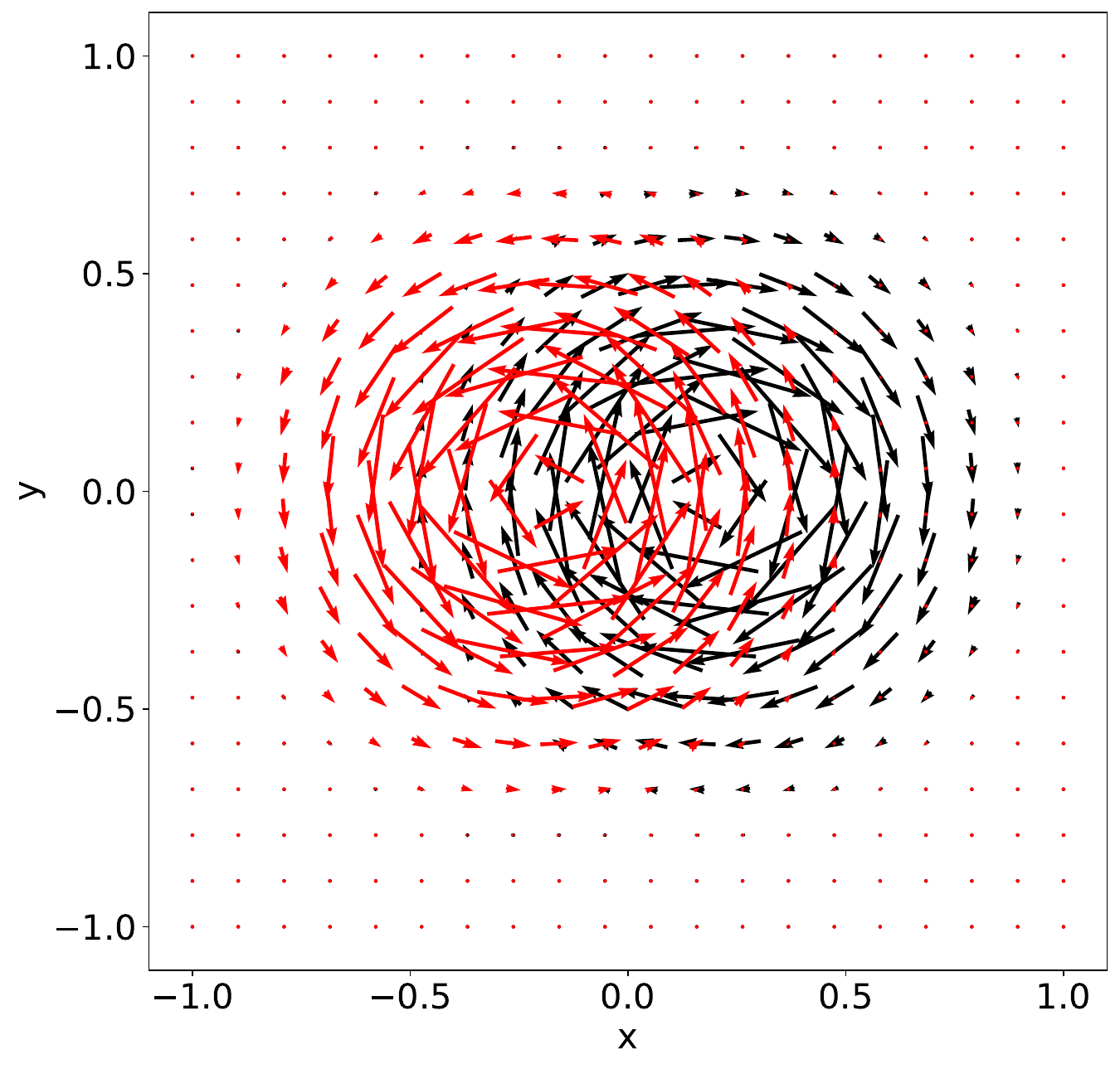}\\
    (b) \includegraphics[width=0.92\linewidth]{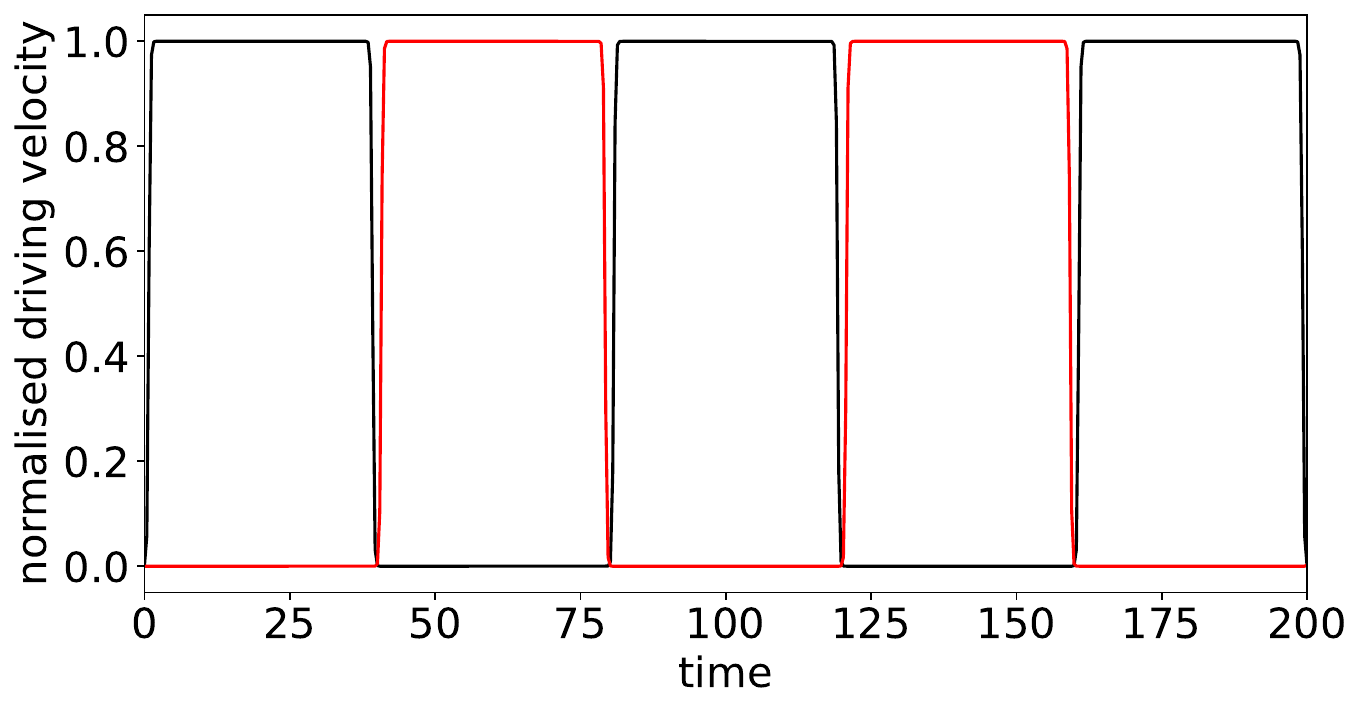}
    \caption{(a) Spatial pattern of the two separate vortex flows applied on the boundary. (b) Temporal variation of the two vortex flows. On {$z=-10$} the red (black)  curve in (b) describes the time variation of the red (black) vortex in (a). On {$z=+10$} the red (black)  curve in (b) describes the time variation of the black (red) vortex in (a).}
    \label{fig:driver}
\end{figure}
The equations are solved in a domain $x,y\in [-1,1]$ and $z\in[-10,10]$. The $x$ and $y$ boundaries are periodic, while for the $z$ boundaries we line-tie the magnetic field while imposing a horizontal driving velocity. This velocity is chosen to be incompressible, and therefore we hold $\rho$, $p$ and $B_z$ fixed, while $B_x$ and $B_y$ are symmetric and $v_z=0$. The tangential driving flow describes two offset counter-rotating vortices that ``blink'' alternately on and off (see Figure \ref{fig:driver}). They inject a net-zero twist into the domain (over a single period of two rotations) and induce a ``pigtail'' braid to a subset of field lines. This is built on a series of previous studies, wherein further details can be found \citep[][]{wilmotsmith2010,pontin2011a,ritchie2016}. The specific form of the boundary driving is given in Appendix \ref{app:driver}. Each boundary rotation executes a maximum of one-half turn, and the peak driving speed is approximately 2\% of the Alfv\'en speed. 

We run three simulations that are identical other than the numerical resolution. The three resolutions chosen are $264^2\times 640$, $528^2\times 1280$ and $1056^2\times 1280$. Due to the use of grid-dependent hyper-resistivity and hyper-viscosity, the three simulations represent three different values of the Reynolds number $R_e\tb{=LV/\nu}$ and magnetic Reynolds number $R_m\tb{=LV/\eta}$ \tb{(where $L$ is a typical length scale, $V$ a typical velocity scale, $\nu$ the kinematic viscosity and $\eta$ the resistivity)}. Since the strongest gradients are in the $xy$-plane, and the resolution is double in this direction between successive simulations, the effective $R_m$ is approximately doubled from the low resolution to the medium resolution simulation, and doubled again at the highest resolution. \tb{Note that the ``effective'' numerical diffusivity is space- and time-dependent, and is difficult to estimate because it depends not only on the algorithm but also on the test case \citep{rembiasz2017}. However, where sharp gradients occur it can be estimated as
$\eta_\mathrm{eff}\sim C\,\Delta x\,c_f$, where $c_f$ is the fast mode speed, $\Delta x$ is the grid spacing, and $C\sim0.1-1$ is a constant that depends on local flow and field gradients \citep{nordlund1997}.}

The Alfv\'en travel time along the domain is 20 non-dimensional units, while each twisting motion takes 40 units.
The low and medium resolution simulations were run until $t=500$ to ensure that the statistically-steady state -- reached after $t\approx 180$ -- is robust at long times. Due to the computational expense, the highest resolution simulation was only run until $t=300$.

\section{Magnetic complexity and energetics} \label{sec:globalresults}
\tb{Our focus in this section is on the global properties of the evolution of the braided field. In Section \ref{sec:sheet_results} we will move to examining the detailed distribution of the current sheets that form and dissipate throughout the evolution.}

\subsection{\tb{Magnetic complexity}}
\label{sec:complexity}

\begin{figure*}
    \centering\includegraphics[width=0.9\linewidth]{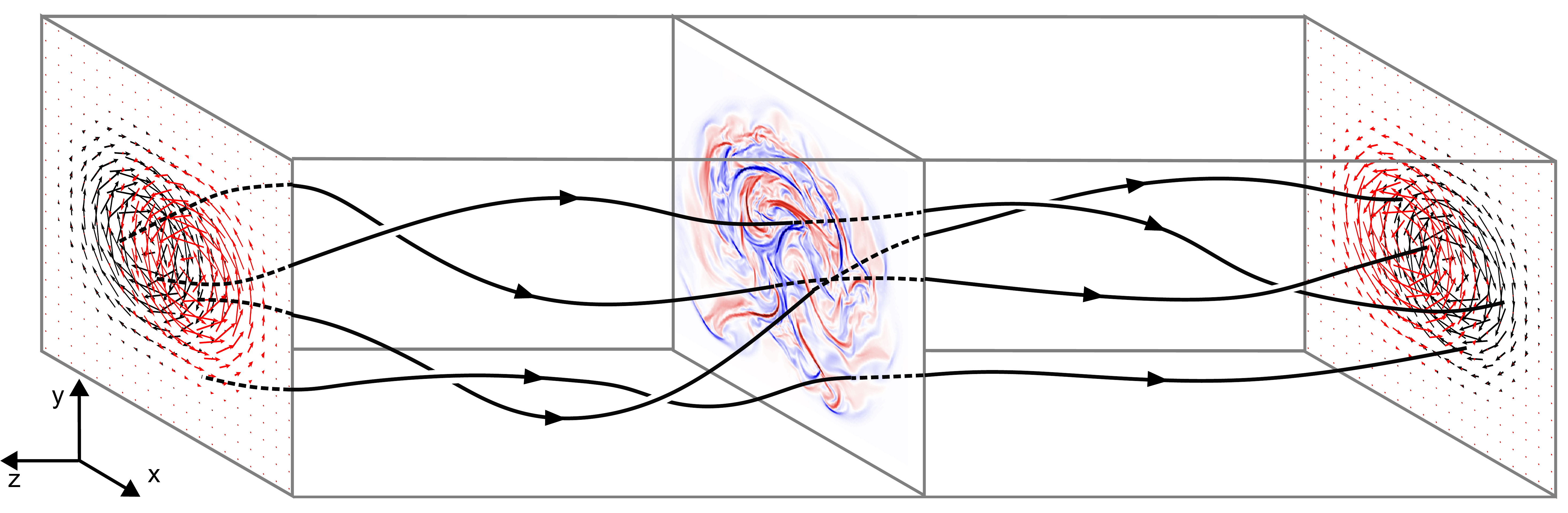}
    \includegraphics[width=0.99\linewidth]{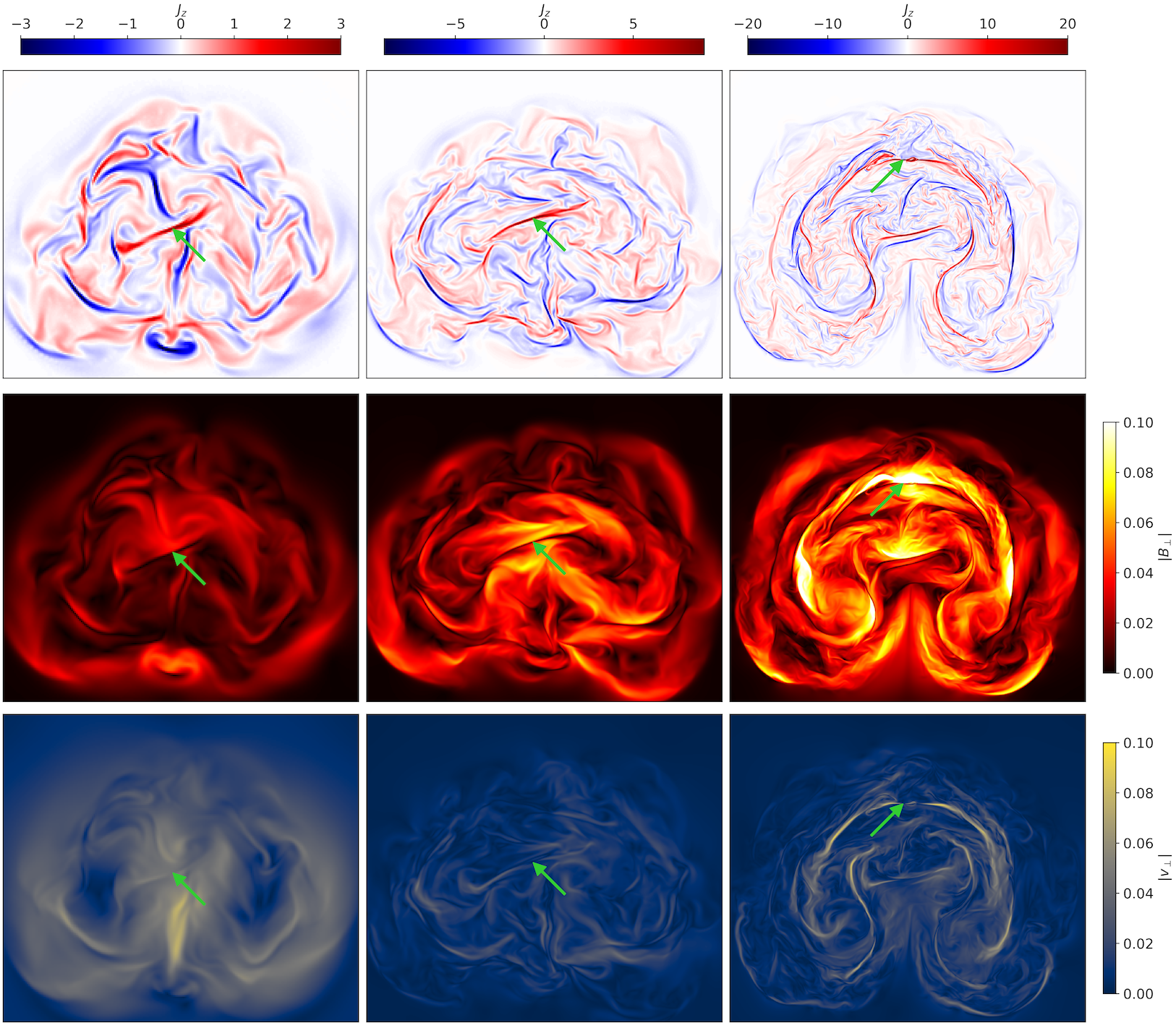}
    \caption{Top: Context image showing the 3D geometry including the driving boundaries and midplane cut. Flow arrows on the two end planes are as in Figure \ref{fig:driver}(a), the colouring of the midplane shows $J_z$, and the lines show schematically the magnetic field line geometry.
    Below that are the distribution in the midplane ($z=0$) of (top to bottom) $J_z$, $|{\bf B}_\perp|$ and $|{\bf v}_\perp|$ for the three simulations at $t=300$, for (left to right) the low, medium and high resolution simulations. The green arrow in each case marks the location of the peak value of $J_z$. {Non-dimensional code units are used for all colour scales.} All frames are plotted over {$x\in [-0.85,0.85]$, $y\in[-0.55,0.92]$.}}
    \label{fig:multipanel}
\end{figure*}

The simulation initial state is an equilibrium, but as soon as the boundary driving is initiated this equilibrium is disturbed, with magnetic energy entering (and in some places leaving) through the driving boundaries. In common with previous flux braiding simulations \citep[reviewed by][]{pontin2020a}, those driving motions twist and tangle the field lines around one another in the volume, leading to the generation of ribbons of current that are highly elongated in the $z$-direction. In much of the following we use a cut through the midplane of the domain in $z$ to analyse these current ribbons/layers -- see the context image at the top of Figure \ref{fig:multipanel}. 

\begin{figure}
    \centering
    \includegraphics[width=0.99\linewidth]{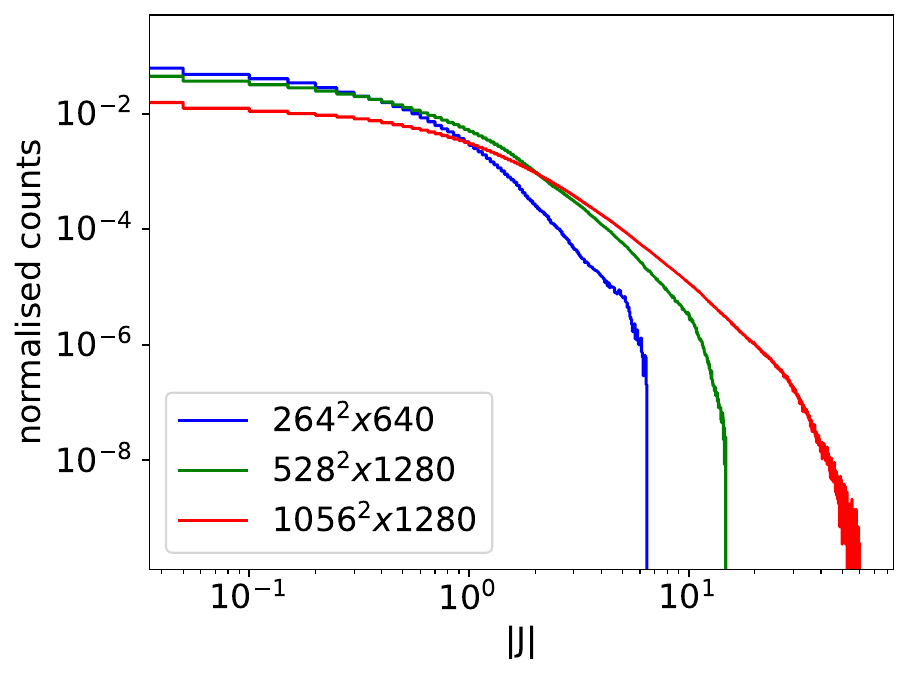}
    \caption{\tb{Normalised distribution of the modulus of the current density at every gridpoint for $x,y\in[-0.8,0.8]$, $z\in[-9,9]$, at $t=300$ for the three simulations.}}
    \label{fig:modj_pdf}
\end{figure}
The lower three rows of Figure \ref{fig:multipanel} show the midplane of the domain ($z=0$) at $t=300$. The \tb{second} row  shows the axial current ($J_z$), from which it is clear that there exists an array of current layers, in which the current flows in both directions. This is expected, due to the fact that the net twist injected by the driver (averaged over space and time) is zero. The other thing that is immediately obvious is that for increasing (magnetic) Reynolds number (resolution) the current layers become thinner, more intense, and more numerous. The details of the current sheet distribution are explored in much more detail in Section \ref{sec:distributions}. \tb{For now we present in Figure \ref{fig:modj_pdf} the current distribution in the 3D volume at $t=300$ for the three simulations, clearly showing higher current values attained for increasing resolution (increasing $R_m$).}

\begin{figure*}
    \centering
    \includegraphics[width=0.99\linewidth]{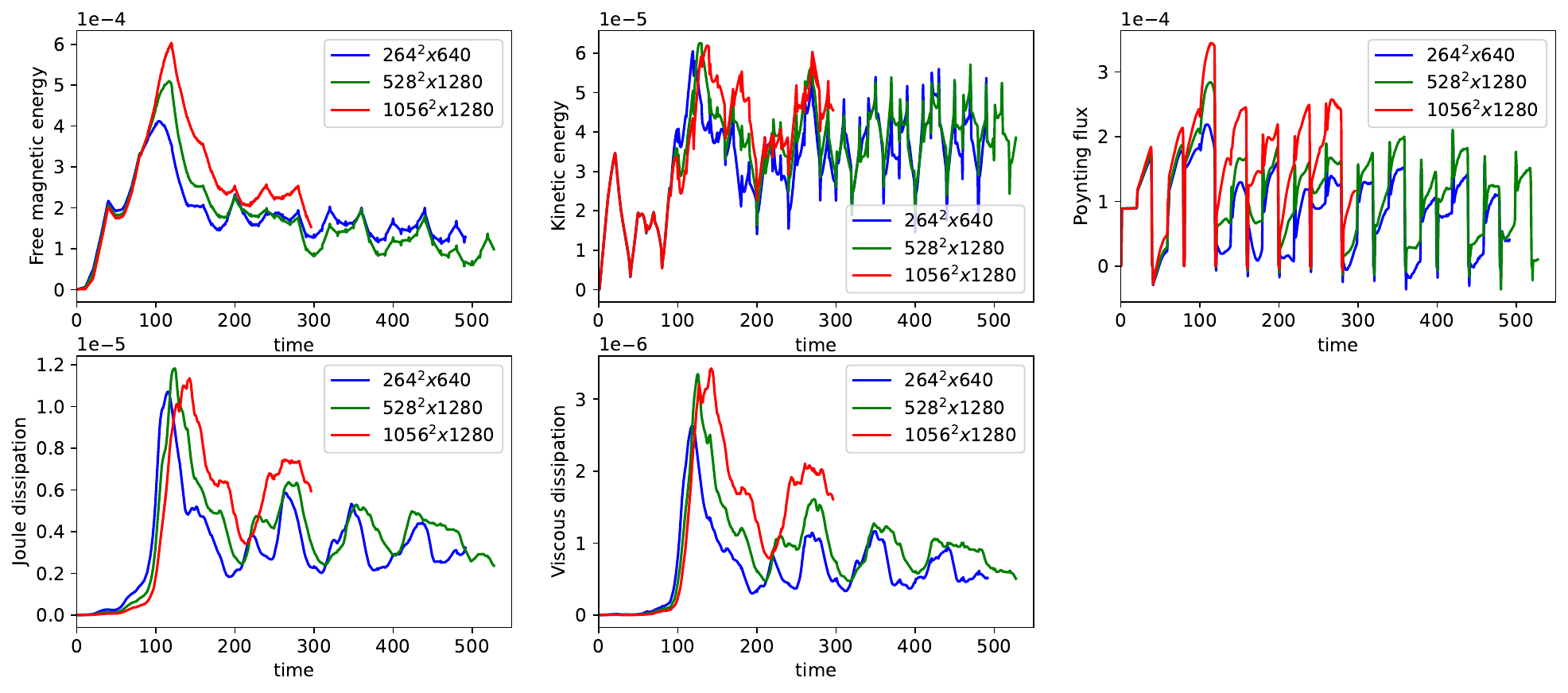}
    \caption{Energetics of the three simulations over time. Top-left to bottom-right: Free magnetic energy (i.e., magnetic energy in excess of the initial state), kinetic energy, Poynting flux through the lower boundary $z=-10$, volume-integrated Joule dissipation, volume-integrated viscous dissipation.}
    \label{fig:energies}
\end{figure*}

The evolution of the volume-integrated magnetic energy -- as well as Joule and viscous dissipation -- reveals more about the global evolution (see Figure \ref{fig:energies}).  All quantities reach a peak at $t\approx 120$ before settling to a statistically steady state after $t\approx 180$. Note that after $t\approx 180$ all quantities show a periodic fluctuation associated with the periodicity of the driver (most clearly seen in the Poynting flux through the lower boundary). 
The fact that the system can store more energy at the early stages is a result of the simple field structure at early time: when the field is simple, the large-scale twisting motion of the driver propagates into the domain and is stored as large-scale magnetic twist within the magnetic field. At later times, the field lines in the domain are tangled, and the stress injected by the large-scale twisting motion on the boundary is distributed across the loop, cascading down to small scales, where it is more easily dissipated. 

\begin{figure*}
    \centering
    \includegraphics[width=0.99\linewidth]{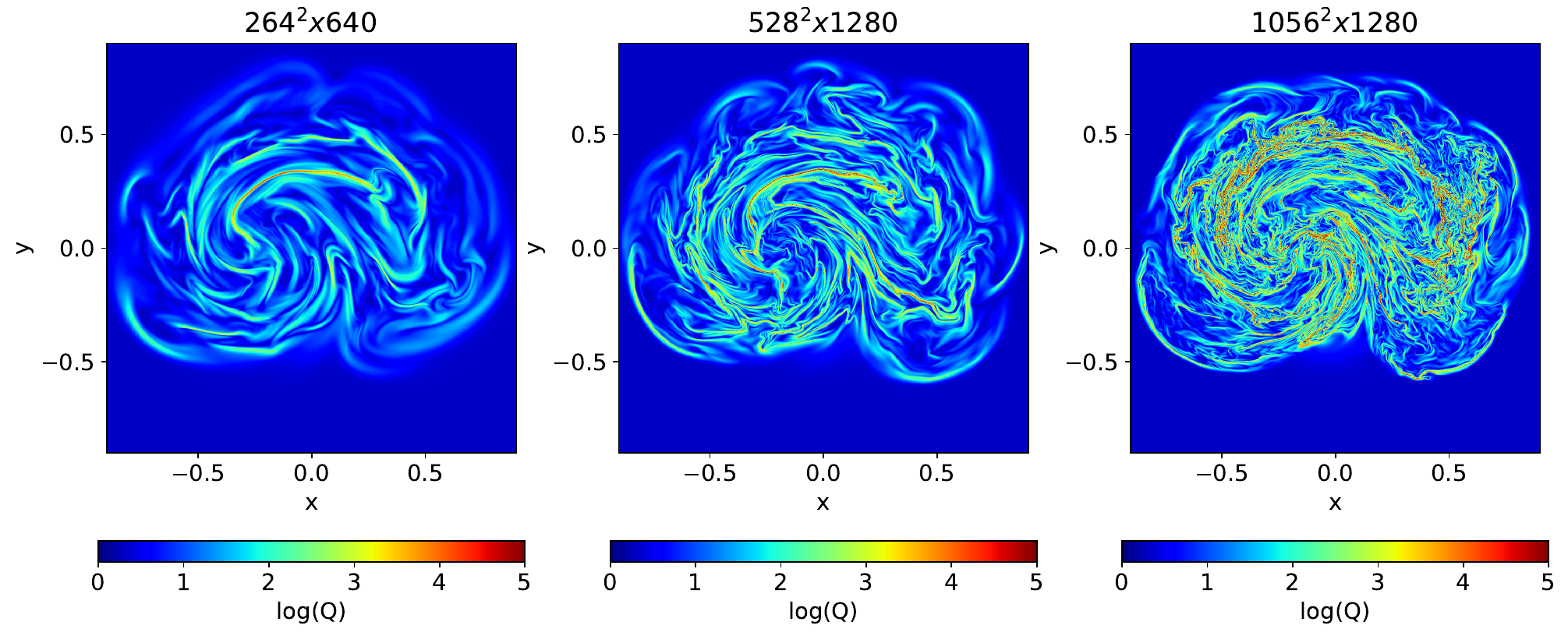}
    \includegraphics[width=0.99\linewidth]{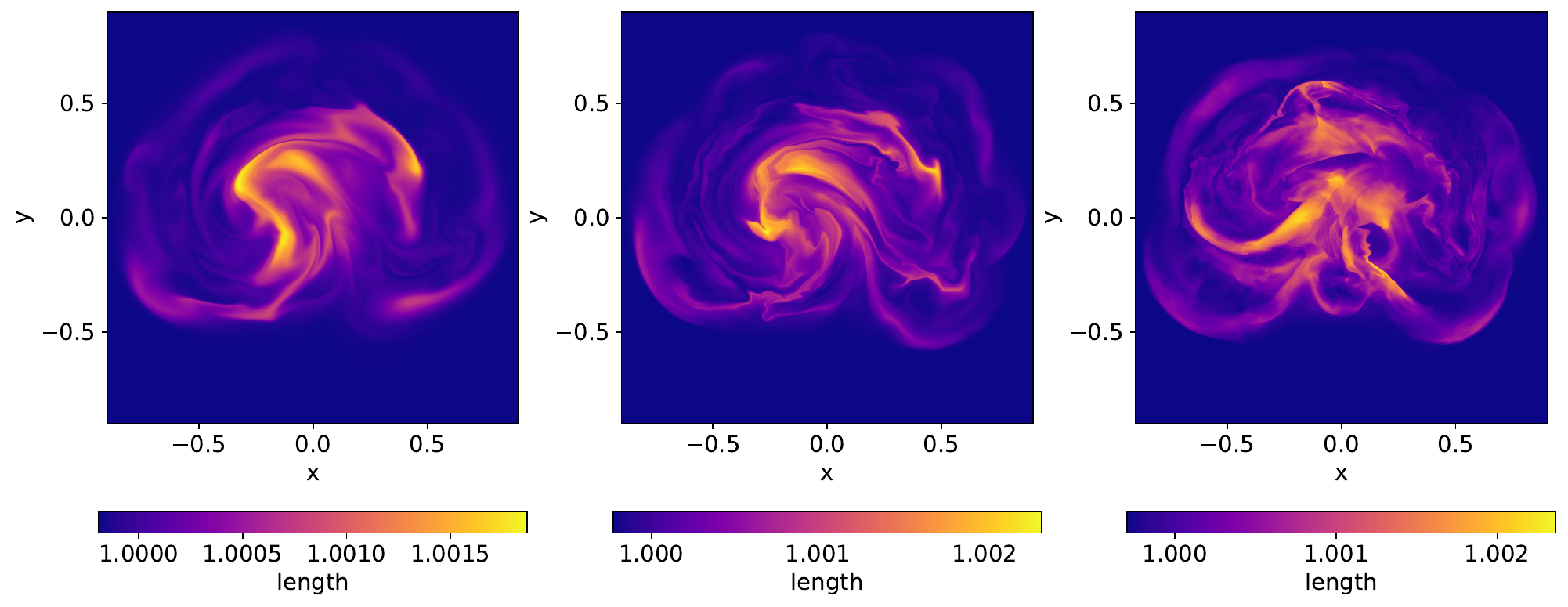}
    
    \caption{Top row: log of the squashing factor, $Q$, plotted on the lower boundary ($z=-10$) at $t=310$ for the three simulations. Bottom row: field line length as a fraction of the initial length, plotted in the midplane ($z=0$) at $t=310$.}
    \label{fig:complexity}
\end{figure*}

The `braiding' or `tangling' or complexity of the magnetic field lines in the domain during the latter part of the simulation can be characterised by calculating the field line mapping from one end of the domain to the other. This can be visualised, for example, by plotting the magnetic squashing factor, $Q$ \citep{titov2007}. We calculate $Q$ using the \texttt{UFiT} package \citep{aslanyan2024} and render it on the lower boundary, $z=-10$, in the top panels of Figure \ref{fig:complexity}. It is immediately clear that as $R_m$ (the resolution) is increased, the state of the field at later time becomes increasingly tangled. This is consistent with the increased number of current layers {at higher $R_M$ (see Table 1)}, and is shown by the higher values and more space-filling nature of $Q$. It is noteworthy that the region of high $Q$ is slightly more compact at higher resolution: we attribute this to the fact that for large $R_m$ it is `more difficult' to initiate reconnection (more stress needs to be built up locally -- see the discussion of ${\bf B}_\perp$ and $B_{\rm in}$ later), and therefore the cross-field spreading (in time) of field line tangling occurs less readily at large $R_m$. {Note that this is consistent with the  delays present in the two dissipation curves before $t = 200$ in Fig.~\ref{fig:energies}.}

A similar phenomenon is observed in the field line length, plotted in the loop mid-plane in the lower panels of Figure \ref{fig:complexity}. Also evident in this plot are the presence at high $R_m$ of local pockets in which the field lines are significantly longer than their neighbours. The peak field line length increases slightly with resolution, but always stays well below a 1\% increase on the initial length, due to the large aspect ratio of the domain.  

\tb{Finally, it is interesting to compare the distribution of current sheets at fixed time between simulations. The second row of Figure \ref{fig:multipanel} reveals that the strong current sheets appear in similar locations for the low-$R_m$ and medium-$R_m$ simulations (at the same times), but in the high-$R_m$ case the correspondence is significantly weaker. It is also the case that if we plot $J_z$ at, say, $z=-5$ and $z=5$ then at low- and medium-$R_m$ the two distributions show distinct similarities, but again this is no longer the case at high-$R_m$. All of this suggests that for the highest $R_m$ there is a transition towards a more chaotic evolution, likely accelerated by current sheet fragmentation -- possibly but not necessarily related to turbulence (see below) that we do not explore further here. As mentioned in Section \ref{sec:setup}, we cannot estimate a single $R_m$ for the simulations, but since the current sheet outflows are on the order of the Alfv\'en speed, $v_A\approx c_f$ then using the loop diameter for $L$ and the effective resistivity from Section \ref{sec:setup} we have $R_m\sim S\sim \frac{L}{C\Delta x}$ which is of order $10^3$--$10^4$ for the highest resolution simulation (where the Lundquist number, $S=Lv_A/\eta$). The larger value could lead to turbulence in an open system, but whether the magnetically closed corona is a turbulent system is under debate \citep{klimchuk2021}.}


\begin{table}
\centering
\begin{tabular}{ l || c c  c }
 simulation resolution & $264^2\!\times\! 640$ & $528^2\!\times\! 1280$ & $1056^2\!\times\! 1280$ \\\hline 
 $\langle$no.~of J sheets$\rangle$ & 12& 23& 26\\
 sheet filling factor & 0.041& 0.029& 0.0095\\
 max($|B_\perp|$) & 0.163& 0.229& 0.260\\

$\langle|B_\perp|\rangle$ & 0.0261& 0.0296& 0.0341\\

$\langle\phi\rangle$ & 1.602$^\circ$ &
1.801$^\circ$ &
2.132$^\circ$\\

max($v_\perp$) & 0.107& 0.146& 0.159\\

$\langle v_\perp\rangle$ & 0.0155& 0.0157& 0.0143\\

$\langle$Poynting flux$\rangle$ & 7.18e-5 & 1.05e-4 & 1.75e-4 \\
\tb{$\langle$Joule dissipation$\rangle$} & 3.82e-6 & 5.17e-6 & 6.76e-6\\
\tb{$\langle$viscous dissipation$\rangle$} & 7.76e-7 & 1.25e-6 & 1.86e-6\\
\tb{$\langle$free magn.~energy$\rangle$} & 1.82e-4 & 1.96e-4 & 2.61e-4\\
\tb{$\langle$kinetic energy$\rangle$} & 3.61e-5 & 3.96e-5 & 4.61e-5
\label{table1}
\end{tabular}
\caption{Mean values \tb{(denoted by $\langle\ldots\rangle$)} of various quantities relevant to the current sheet distribution and energetics, \tb{all calculated for $125\leq t\leq 300$, i.e.~the period over which data exists for all three simulations}. Mean values of $|B_\perp|$, $|v_\perp|$ {and the magnetic vector angle $\phi=\arctan(|{\bf B}_\perp|/B_z)$ are calculated over {$x,y\in[-0.5,0.5]$, $z\in[-10,10]$. \tb{Other quantities are calculated for the full domain.}}}}
\end{table}

\subsection{\tb{Energetics and heating}}\label{sec:energy}

\tb{To examine the effect of $R_m$ on the overall energy release process we evaluate the mean and maximum values of various quantities found at later simulation times ($125\leq t\leq 300$) -- see Table 1.}
Owing to the simplified thermodynamics and plasma structure in our simulations, we cannot say anything directly applicable to coronal observations about loop temperatures or morphologies. Nevertheless, our results provide some important insights for more sophisticated models that, for computational necessity, are not able to access the range of (magnetic) Reynolds numbers that we have explored here.

First, we calculate the filling factor of current sheets across field lines in the volume. This is approximated by simply calculating the filling factor of pixels identified as being within current sheets by the algorithm described in Appendix \ref{app:edge}, averaged over time from $t=125$ onwards. The results are given in Table 1. There are two competing effects: current sheets get thinner as $R_m$ increases, but also more numerous. The results show that the net effect is that the filling factor decreases as $R_m$ increases.

While the filling factor of current sheets decreases with increasing $R_m$, the associated energy is still expected to be spread broadly across the loop. We find the highest temperatures in the current sheets and their immediate outflows. But the temperature is moderately high within the whole tube, as the flux that has been processed through the current layers gets rearranged across the loop.  This redistribution of the heat across the loop will be aided in the corona by the efficient field-aligned thermal conduction, and the presence of tangled field lines provides an effective 
``cross-field'' energy transport, where here we refer to transport across the mean or guide field (in this case, $B_z$). As described in Section \ref{sec:complexity} and shown in Figure \ref{fig:complexity}, this field line tangling is significantly enhanced at larger $R_m$. The same argument applies to the upflow of material due to evaporation.

One possible source of increasing field line tangling would be an increase in $|{\bf B}_\perp|$ -- i.e., $|{\bf B}_{xy}|$. 
The mean $|{\bf B}_\perp|$ in the vicinity of the driving boundaries is also critical in determining the Poynting flux injected by the driving flow. The maximum and mean values for $t\geq 125$ are shown in Table 1 (for $x,y\in[-0.5,0.5]$)\tb{, as well as the free magnetic, which is energy associated with this component of the field. All of these quantities} are seen to increase as $R_m$ increases, with the mean value of $|{\bf B}_\perp|$ increasing by $\approx 30\%$ from the lowest to highest resolution simulation, while the maximum value increases by $\approx 60\%$. {As argued by \cite{parker1988}, the mean $|{\bf B}_\perp|$ must be sufficiently large to ensure that the Poynting flux into the loop can balance the radiative losses.}
Examining the Poynting flux through the lower driven boundary (Figure \ref{fig:energies}, top-right panel and Table 1) we observe a corresponding increase with $R_m$. {As shown in Table 1, the mean angle of the magnetic field vector to the vertical, $\phi$, is somewhat below the $14^\circ$ value predicted by \cite{parker1988}. However, it is increasing with $R_m$, and it should be noted that Parker's estimate was based on a field uniformly sheared along its length.}

Finally, we comment on the nature of the energy dissipation in the domain. 
Here we have used similar rules for the grid-dependent hyper-resistivity and hyper-viscosity, as described in Section \ref{sec:setup}. We find (see Figure \ref{fig:energies} lower panel \tb{and Table 1}) that the Joule dissipation is larger than the viscous dissipation by a factor of 2-3. We note, however, that in the corona the fluid Reynolds number is thought to be orders of magnitude larger than $R_m$, so that viscous damping of reconnection outflow jets may dominate on the Sun. 

\section{Distribution of current sheets}\label{sec:sheet_results}

\begin{figure*}
    \centering
(a)    \includegraphics[width=0.46\linewidth]{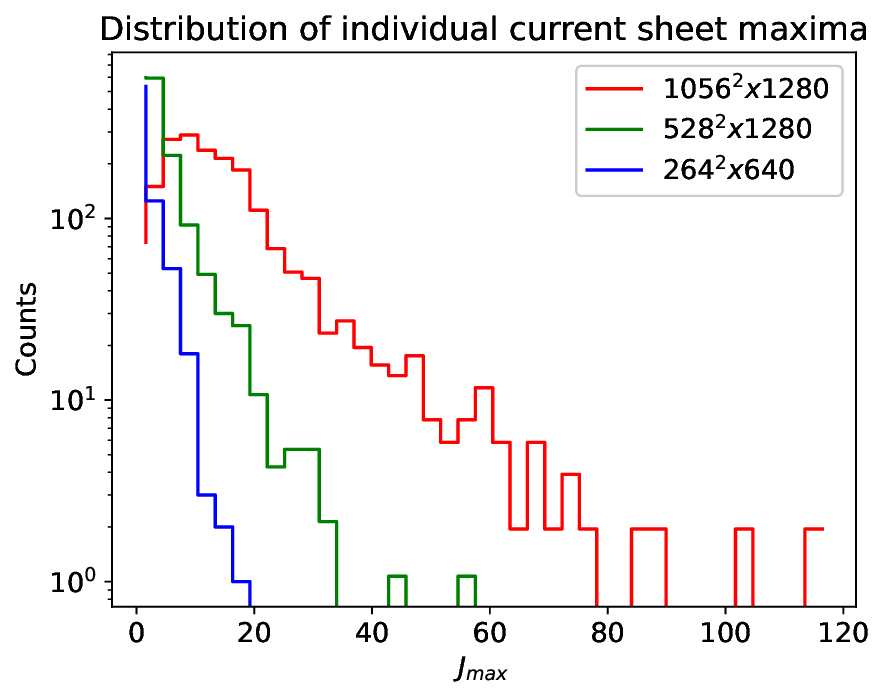}
(b) \includegraphics[width=0.46\linewidth]{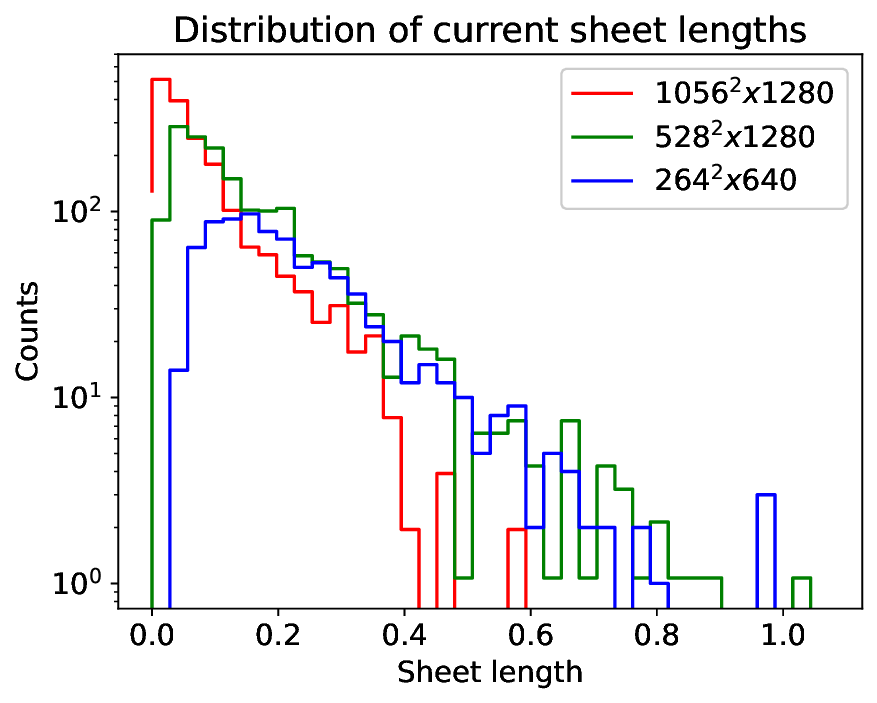}\\
(c)    \includegraphics[width=0.46\linewidth]{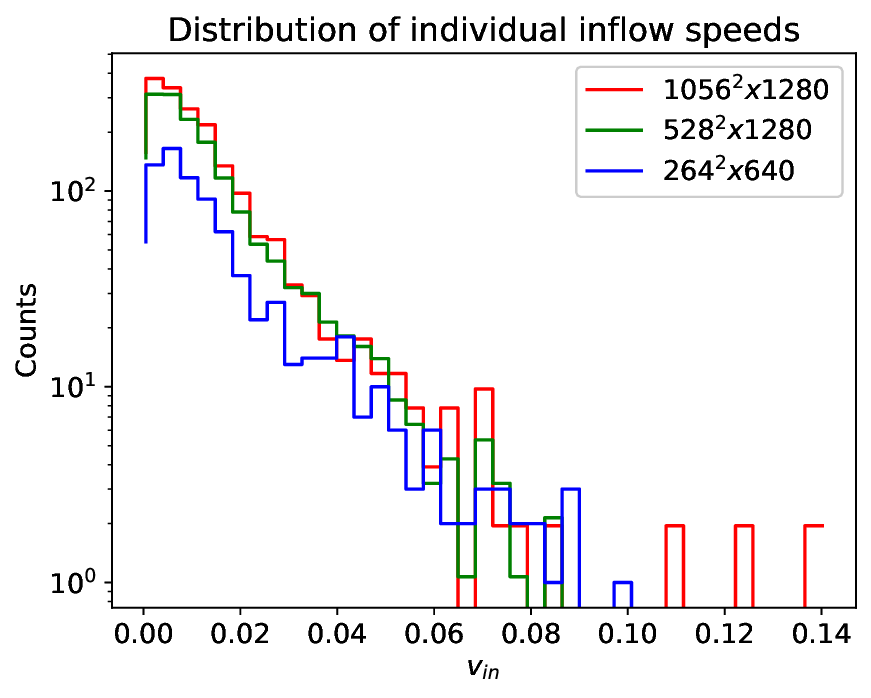}
(d)    \includegraphics[width=0.46\linewidth]{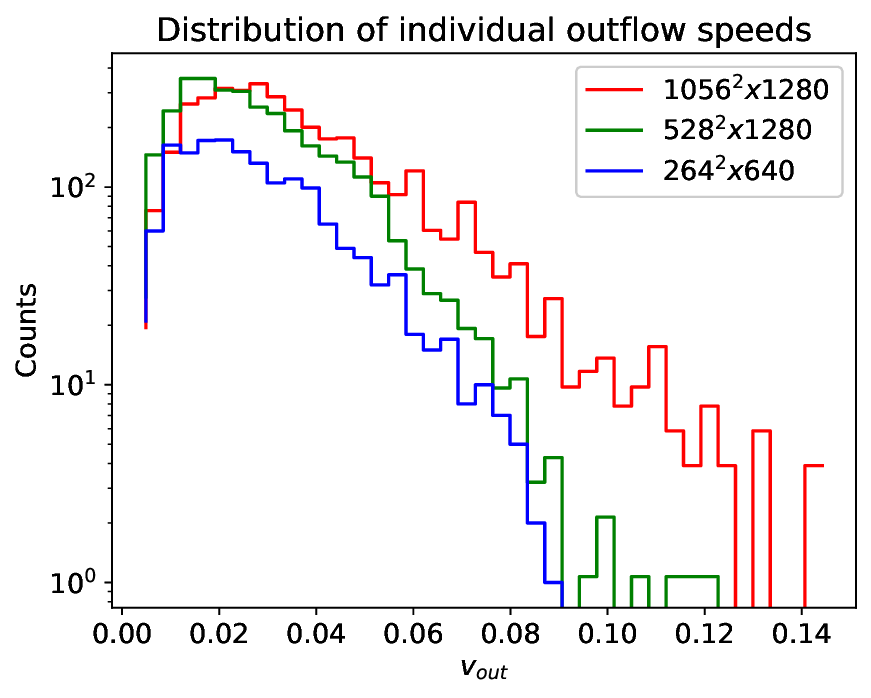}\\
(e)    \includegraphics[width=0.46\linewidth]{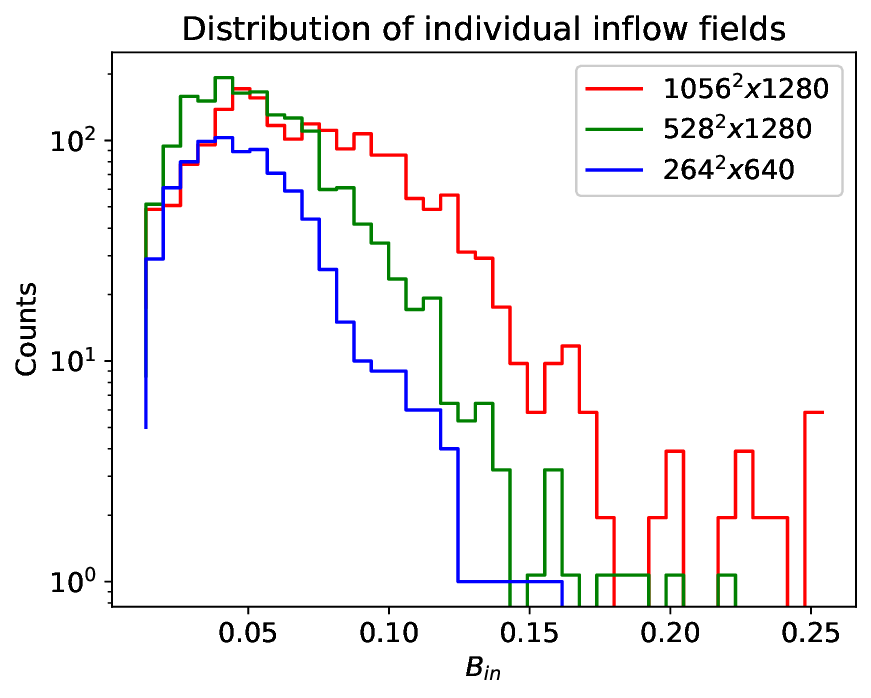}
(f)    \includegraphics[width=0.46\linewidth]{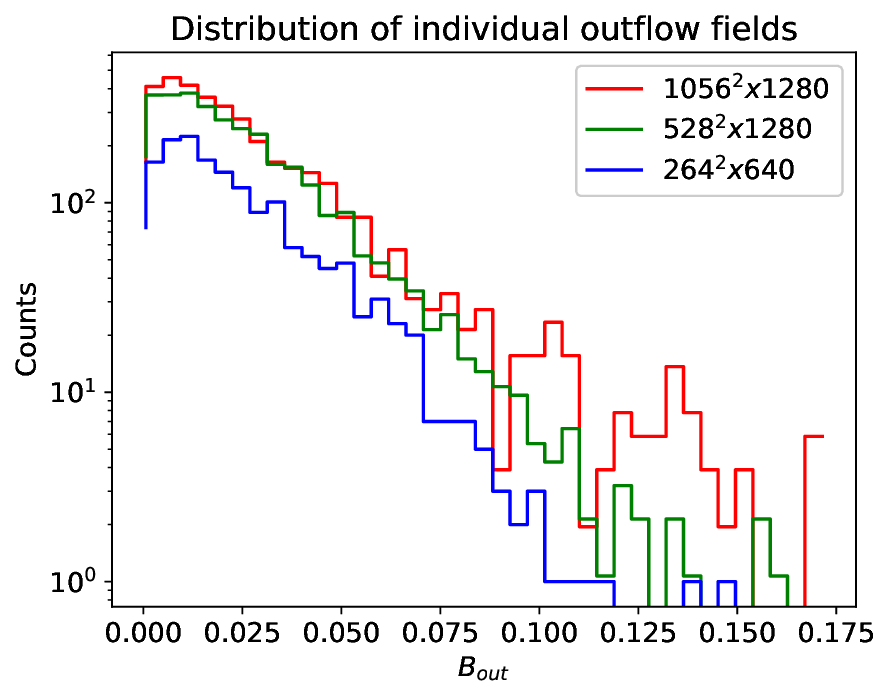}
    \caption{Distributions of current sheet properties \tb{measured in the mid-plane ($z=0$)} for the three different simulations, \tb{for all time}, calculated as described in Appendix \ref{app:edge}. Note that in (b), `length' refers to the length in the $z=0$ plane, \tb{and in all cases non-dimensional code units are used. For reference, the Alfv\'en speed based on $B_z$ is $v_A\approx 1.$}}
    \label{fig:distributions}
\end{figure*}

\subsection{Methodology for current sheet identification}
As shown in the top panels of Figure \ref{fig:multipanel}, the current density is substantial within a significant part of the domain cross-section. (Note that this axial ($z$) component of ${\bf J}$ is larger by a factor of typically 20 than the in-plane ($xy$) components.) Given the number of current sheets at even a single time, an automated identification method is necessary. Many such feature identification algorithms exist for various applications in astrophysics \tb{-- an example of usage in solar physics is \cite[e.g.,][]{deforest2007}}. 

Here we employ an approach that is specific to the morphology of current sheets. Specifically, a current sheet has a high aspect ratio, and if it is actively reconnecting then it will be only a few grid cells across (due to the hyper-resistivity employed). 
As a result, current sheets tend to exhibit sharp `edges' (in the current density distribution) and so we start by identifying such edges. This has the advantage over a traditional clumping algorithm that it preferentially identifies thin, sheet-like structures. It also has the advantage over a simple contour/threshold definition that is can allow us to identify the many small, less intense sheets present particularly in the highest resolution runs -- i.e., it is more suited to identifying structures with a hierarchy of scales and intensities. 

\tb{In this study we quantify the properties of the current sheets by identifying them in the $z=0$ plane (what would be the ``loop apex'' in a curved geometry). The simplification allows for application of established image processing techniques and comparison with 2D reconnection models. 
This allows a representative comparison of current sheets between the different simulations, on the basis that from visual inspection most current sheets pass through the mid-plane, and have their long axis almost perpendicular to the mid-plane. To quantify the latter, we find the location of the peak current density for each identified current sheet and at that location calculate the angle between the current vector and the $z$-axis.
For example, in the $528^2\times1280$ simulation both the mean and median angle are $1.5^\circ$.
An extension to an analysis of the current sheets as fully 3D objects should be done in future.}

The edge detection algorithm that we use \tb{to identify current sheets in the $z=0$ plane} is described in detail in Appendix \ref{app:edge}. Having run it on  every snapshot of each simulation, the properties of the individual sheets as well as the magnetic field and velocity profiles in their vicinity can be measured (see Appendix \ref{app:edge}). The results are presented next. 

\subsection{Properties of the distribution of current sheets}\label{sec:distributions}

\begin{figure*}
    \centering
   (a)\includegraphics[width=0.57\linewidth]{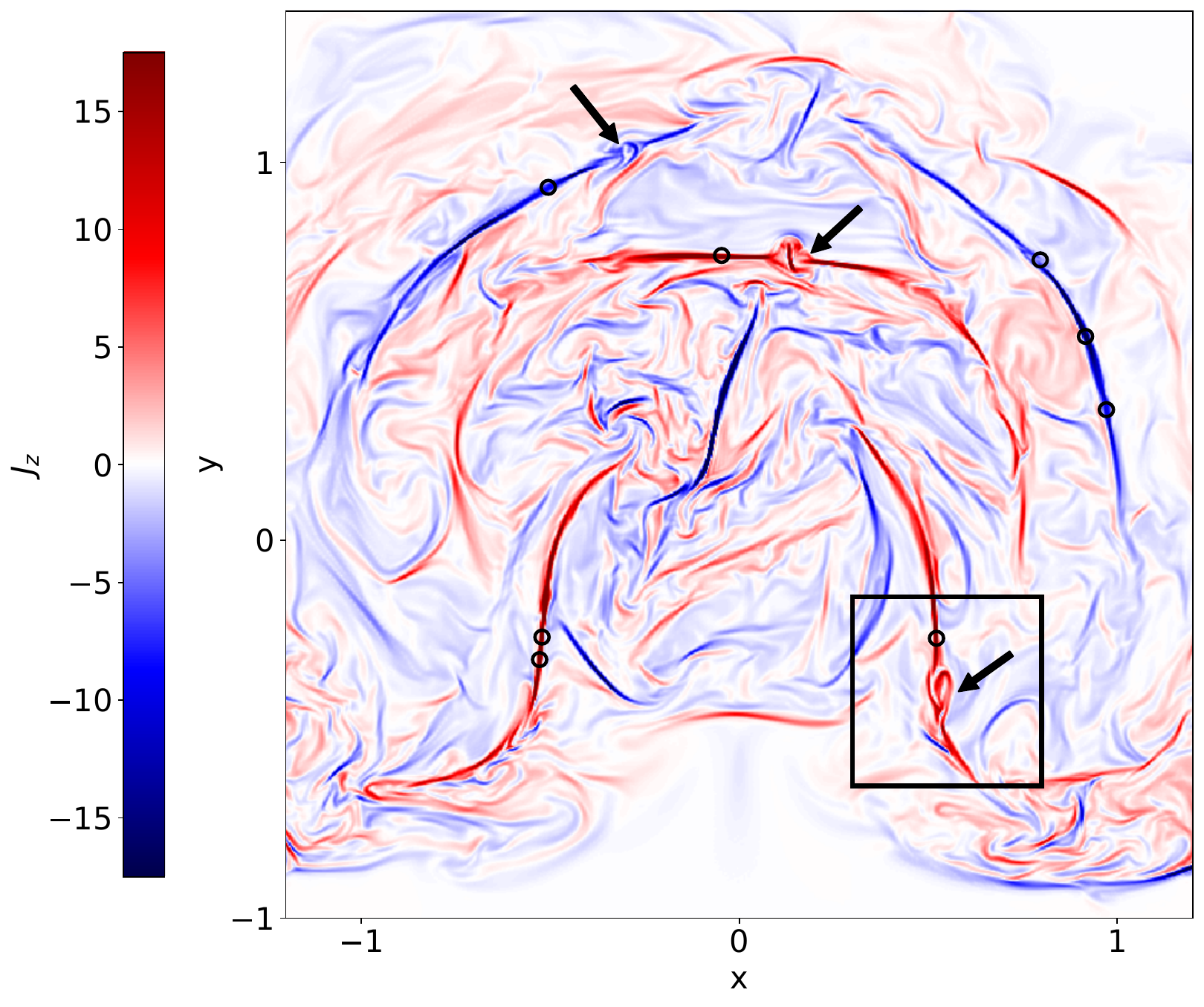}
   (b)\includegraphics[width=0.37\linewidth]{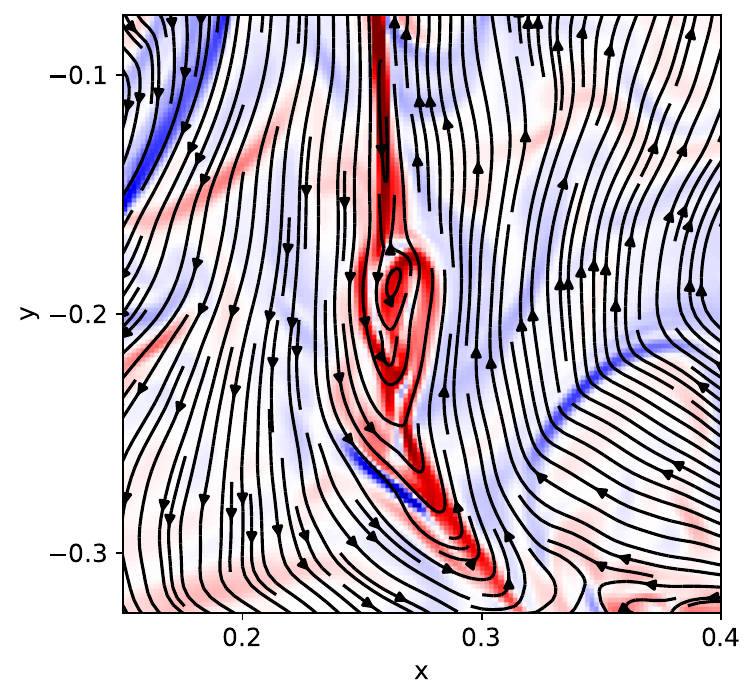} 
   \caption{Distribution of $J_z$ in the midplane at $t=270$ for the simulation with resolution $1056^2\times 1280$. There are a number of locations in which a current sheet appears fragmented along its length with bubble-like features suggesting the presence of plasmoids and/or secondary current sheets. In (a) the most prominent are marked with arrows, while the locations of additional, smaller O-type structures within current layers are marked with black circles. (b) shows a close-up of the region inside the square in part (a) including some sample magnetic streamlines.}
    \label{fig:plasmoid}
\end{figure*}

The statistics of the current sheets and their properties are plotted in histograms (with logarithmic $y$-axes) in Figure \ref{fig:distributions}, from which we observe the following. First (Figure \ref{fig:distributions}a), as expected, for higher $R_m$ the current sheets tend to be more intense -- the typical maximum value of $|{\bf J}|$ approximately doubles when the resolution is doubled. 
Figure \ref{fig:distributions}b shows {the distribution of the length of the current sheets, measured in the $z=0$ plane. We observe} that for moderate sheet lengths (0.1--0.3 units) the distribution of sheet lengths is similar between the three simulations {(though shorter sheets can naturally be detected at higher resolutions)}. This trend continues for the lower two resolution simulations up to sheet length of $\sim 0.8$. However, for the highest resolution simulation there appears to be a rather hard cut-off limit for sheet length at around 0.5. We propose that this is because we have accessed a new regime where current sheets exhibit tearing, breaking into shorter sheets separated by plasmoids. This is based on (i) visual inspection of the current distribution -- see, for example, Figure \ref{fig:plasmoid} -- and (ii) the current sheet aspect ratio. 
\tb{The onset of the plasmoid instability typically requires both the Lundquist number to exceed $\sim10^4$ and the current sheet aspect ratio to exceed $\sim50-100$ \citep{biskamp1986,ni2010,loureiro2013,huang2013}.
To estimate the latter in our simulations}, note that the current sheet thickness (for all simulations) is on the order of 6-7 pixels, and for the highest resolution run this corresponds to a \tb{thickness} of $\sim 0.011$. Therefore a current sheet length of 0.5 would give an aspect ratio of $\sim 50$, \tb{consistent with plasmoid instability onset based on the above conditions. By contrast, for the medium-$R_m$ simulation the minimum current sheet thickness is $\sim 0.022$, so a sheet aspect ratio of 50 would require a sheet length of $\sim$1 -- }
Note that while the above thresholds typically come from studies in a strictly 2D geometry, a similar threshold has been demonstrated for fully-3D current sheets \citep{wyper2014a}, and the dependence of tearing on the sheet length in a geometry like the one considered here has been explored by \cite{leake2024}. \tb{For further discussion see Section \ref{sec:sub:tearing}}.

Figure \ref{fig:distributions}(c-f) show distributions of the current sheet inflow and outflow speeds and field strengths (with notation as per the classic Sweet-Parker model), calculated as per Appendix \ref{app:edge}. First, we see that the inflow speeds 
{depend (at most) weakly on}
$R_m$. On the other hand, outflow speeds are {systematically} 
larger for larger $R_m$, which is expected because the larger currents will generate larger Lorentz forces to drive the outflow. {The higher outflow speed can also be considered a consequence of the mass conservation in the sheet: the inflow speed (and thus mass flux) is insensitive to $R_m$ (across the distribution), and since the current sheets are thinner, the outflow must be faster to provide the outflow mass flux to balance the inflow.}
Both the inflow and (to a lesser extent) outflow field strengths show a tendency to increase with $R_m$, though the tendency is quite weak for $B_{\rm out}$. This is consistent with the overall increase in the peak and mean values of $|{\bf B}_{xy}|$, see Section \ref{sec:energy}.

\subsection{Comparing current sheet properties with reconnection models}
\subsubsection{Comparison with Sweet-Parker model}

Using the current sheet properties described above we can compare with expectations from theoretical models. The first natural step is to compare $v_{\rm in}, v_{\rm out}, B_{\rm in}, B_{\rm out}$ with the classic Sweet-Parker scaling relations \citep[see, e.g.,][]{pontin2022}. For example, \tb{magnetic flux conservation for a Sweet-Parker current sheet} predicts that $v_{\rm in}B_{\rm out}/v_{\rm out}B_{\rm in}=1$. Evaluating this ratio for our identified current sheets we find that it is usually very different from 1, often by an order of magnitude or more. Some possible reasons for this departure include the fact that the current sheets are not quasi-steady, that we are considering only a 2D cut through a 3D structure, and the difficulty in assigning values to these quantities (see Appendix \ref{app:edge}). Specifically, the current sheets are not well separated (so that their in/outflows influence one another) and they are quite inhomogeneous both along their length and from one side of the current sheet to the other -- therefore the inflow/outflow speeds and field strengths are quite variable depending on exactly where they are measured.
Notably, the values of $|{\bf B}_\perp|$ (i.e., $B_{\rm in}$) on either side of the current sheet can be significantly different, as evident from Figure \ref{fig:multipanel}. \tb{(As noted in Appendix \ref{app:edge}, the value of $|B_\perp|$ assigned to a sheet is different from the average of $|B_\perp|$ on either side of the sheet.)} Furthermore, there are clearly identifiable reconnection outflow `jets' for some current sheets but not all (e.g., Figure \ref{fig:multipanel}). {Comparison of the values of $|{\bf v}_\perp|$ and $|{\bf B}_\perp|$ shows that the jet speeds are comparable to the inflow Alfv\'en speed based on this in-plane field component as predicted by Sweet-Parker theory  (since $\rho\sim1$ throughout the domain.)}

{Disentangling the reasons for the departures from the Sweet-Parker scalings, or obtaining modified scalings, requires a careful analysis of the evolution of a handful of sheets in three dimensions, not just in a horizontal cross-section as done here. We plan to do this in a follow-up study.}

\subsubsection{Current sheet loss of equilibrium}\label{sec:lossofeq}
\cite{klimchuk2023} have argued that for current sheets in a 3D, line-tied geometry 
such as the one considered here, a finite-thickness equilibrium is possible as long as the sheet is not too short. The critical length ($\lambda$) depends on the ratio of the perpendicular to guide field strengths (the shear across the sheet) and the length of the field ($L$) in the guide (line-tied) direction: 
\begin{equation}
    \lambda<\frac{L}{2}\frac{B_{\perp 0}}{B_{z0}}.
\end{equation}
When the inequality is first satisfied -- usually because of increasing shear -- the sheet collapses to a small thickness and starts to reconnect. 
Using the data from Table 1 and Figure \ref{fig:multipanel} we have $L\approx 20$, $B_{z0}\approx 1$, and $B_{\perp 0}\approx 0.03$, which gives $\lambda<0.3$. Examining the distribution of current sheet lengths in Figure \ref{fig:distributions}(b) we would therefore expect the majority of current sheets in the domain to be undergoing ``active'' reconnection (which occurs in practice here when the sheet thickness drops to a few grid cells as determined by the hyper-resistivity model, as discussed above). \tb{Sheets longer than 0.3 will not have experienced a loss-of-equilibrium collapse. Whether they are actively reconnecting depends on whether their equilibrium thickness is greater than or less than a few grid cells.} 

\subsubsection{Secondary instability/non-linear tearing}\label{sec:sub:tearing}

Current sheet fragmentation in 3D, line-tied fields has been observed before, for example in sheared fields \citep{leake2020} and in the context of the coalescence instability \citep{huang2016}. However, to our knowledge this is the first time that it has been observed in the evolving, tangled fields of a flux braiding simulation. We have identified numerous flux ropes within current layers in our highest resolution simulation by visual inspection. A lesser number can also be found at certain times in the medium-resolution simulation, and none at low resolution. 

In some cases the flux ropes appear as large `bubble' features within a current layer, as in the examples marked by arrows in Figure \ref{fig:plasmoid}. In other cases, the entire O-type/spiral of the in-plane field is a thinner, elongated structure contained within the current sheet, with the locations of such features marked by black circles in Figure \ref{fig:plasmoid}. One should treat this preliminary analysis with caution: the presence/location of these elliptic points is dependent on the plane in which the field vectors are plotted. Here we have used a horizontal plane, for convenience, while a more appropriate choice may be, for example, a plane that is perpendicular to the ${\bf B}$ or ${\bf J}$ vector at the location of the current maximum of the sheet. 

{As mentioned in Section \ref{sec:distributions}, the tearing appears to limit the current sheets to an aspect ratio of $\sim50$, consistent with plasmoid instability theory. This in is spite of the fact that the small number of gridpoints across any given current sheet means that the internal tearing layer is totally unresolved. This is consistent with other 3D tearing simulations \citep[e.g.,][]{wyper2014a}.}
As discussed in the introduction, the growth of plasmoid/flux rope structures within the current layer is typically thought to occur in the non-linear phase of tearing (either within the plasmoid instability or secondary instability picture).
The presence of current sheet fragmentation is crucial, in that it offers a route to speed up the reconnection process at coronal $R_m$, and to explain the ``switch-on'' nature of the heating, which is required to explain the coronal energy balance \citep[e.g.][]{klimchuk2015}.  It is clear from Figure \ref{fig:distributions}(b) that the tearing is effective in restricting the current sheet length. If the individual sheets were close to Sweet-Parker-like, this would correspond to an increase in the reconnection rate, but as seen above these Sweet-Parker relations do not hold here. In the future, it will therefore be important to perform a more detailed analysis to determine what can be learned about the applicability of the previous theory and simulation results to our evolving, fully three-dimensional current sheets.

\subsection{Reconnection rate}

\begin{figure}
    \centering
   \includegraphics[width=0.97\linewidth]{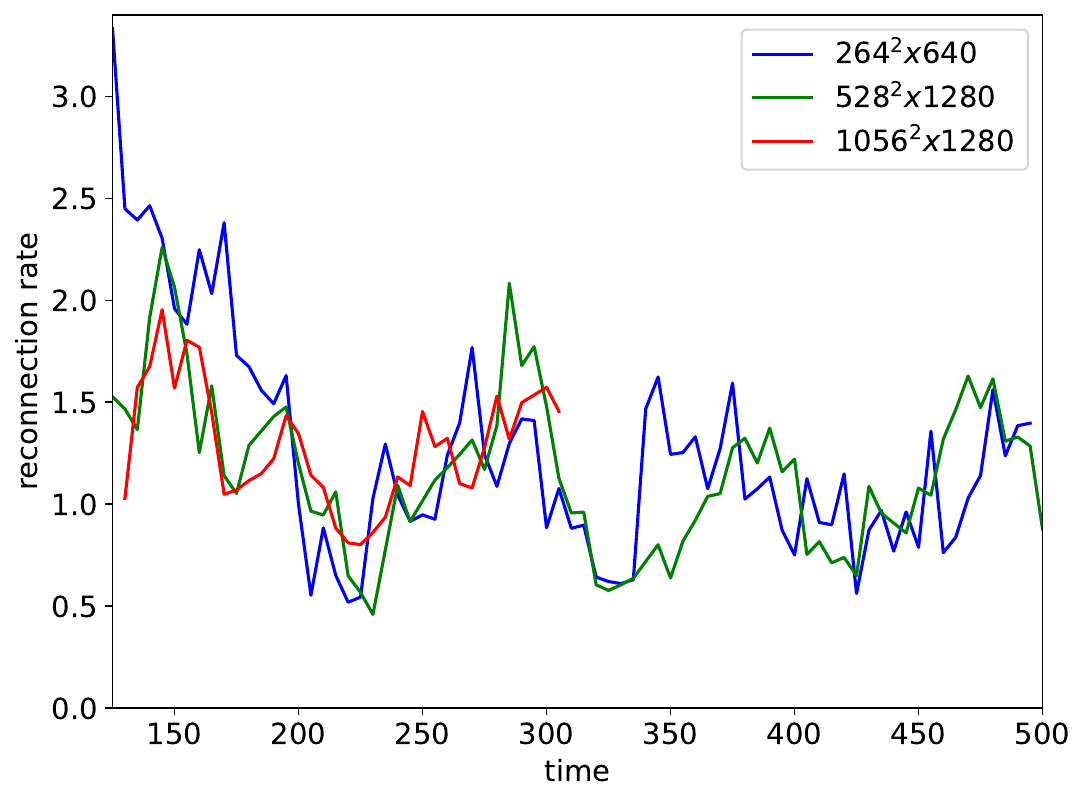}
   \caption{Global reconnection rate as a function of time, calculated as described in Appendix \ref{app:recrate}.}
    \label{fig:recrate}
\end{figure}

An important consideration with any reconnection-mediated energy transfer mechanism is the scaling of the rate of reconnection with $R_m$, due to the extreme value of $R_m$ found in the corona (far out of reach of present day simulations). In 3D, for an isolated reconnection region the reconnection rate is given by the maximal value of the integral of $E_\|={\bf E}\cdot{\bf B}/|{\bf B}|$ over all field lines threading the current sheet \citep{schindler1988,pontin2022}. We note that in our case the relevant reconnection rate is a \emph{net} rate across a fragmented current distribution, where the individual reconnection rates for each current sheet need to be combined \citep{wyper2015}. 

Following \cite{pontin2011a} we estimate the reconnection rate at each time by undertaking the following steps: (i) integrate $E_\|$ along a set of field lines with seed points from a regular grid in the midplane $z=0$ ($800^2$ points equally spaced for {$x,y\in[-0.8,0.8]$}); (ii) identify the local maxima in the absolute value of this 2D distribution; (iii) sum these maxima to get a global reconnection rate produced by the net effect of all of the individual reconnection processes. The resultant reconnection rate depends on the procedure used to identify the local maxima to sum -- we use the same procedure as that used for identifying current sheets, as discussed in Appendix \ref{app:recrate}. However, regardless of the choice of parameters, we find that the global reconnection rate is insensitive to $R_m$ -- see Figure \ref{fig:recrate}. This is consistent with the results found for braid relaxation experiments (i.e., in the absence of boundary driving) by \cite{pontin2011a}. {Note that by contrast the Poynting flux into the domain and the energy dissipation rate both increase with $R_m$ -- see Table 1.}

{It is worth pausing to consider what fraction of the current sheets in the domain are ``actively reconnecting". Due to the hyper-resistivity used, $E_\|$ becomes significant when the sheet thickness reaches $\sim6-8$ pixels. Examining the distribution of $J_z$ for a given simulation run (for example in Fig.~\ref{fig:multipanel} or \ref{fig:algorithm}) we observe, by eye, a relatively uniform thickness (corresponding to $6-8$ pixels).  Our algorithm is designed to find sharp edges, and therefore preferentially finds the thin sheets, though some broader, more diffuse current structures do exist  (e.g., bottom-right quadrant of Fig 8), but are not identified. We should expect to see sheets in the formation stage, when they might be broader, before eventually thinning down to the dissipation scale {(perhaps due to the loss of equilibrium mechanism described in Section \ref{sec:lossofeq})}. 
Note, however, that the time resolution of our simulation outputs is low, and the fact that most of the current structures that we see seem to be down at the ‘reconnecting scale’ of a few pixels, suggests {either (i) that the sheets are only detected by the algorithm when they thin to the grid scale following loss of equilibrium, and/or (ii)} that the thinning/formation time is short compared with the reconnecting/dissipating time.  This dissipation time for individual sheets is likely be artificially long compared to what would be found on the Sun, because the current sheets are laminar, while the plasmoid instability sets in only in the highest resolution simulation.} {We would not expect such laminar (``Sweet-Parker-like'') current layers to be reconnecting for coronal parameters -- they would reach the onset threshold for rapid tearing (possibly preceded by a loss of equilibrium and associated rapid thinning) long before this occurred.}

\section{Conclusion}\label{sec:conc}
Energy conversion 
\tb{mediated by braiding-driven reconnection} 
is certain to play some role in explaining the heating of the solar chromosphere and corona
\tb{(see discussion and references in Section \ref{sec:intro})}. 
One of the great challenges in assessing the role that this mechanism plays in coronal heating has always been the uncertainty in extrapolating simulation results from numerically accessible parameter regimes to solar parameters. To this end, in this paper we have undertaken to study the scaling of various properties of the energy release process with the fluid and magnetic Reynolds numbers. We summarise our results as follows:
\begin{enumerate}
\item 
For increasing $R_m$, the current sheets become thinner, more intense, and more numerous.
\item 
Using a combination of edge detection and region labelling algorithms we can automate the detection of current sheets and analyse their properties. Current densities and inflow and outflow speeds and field strength show approximately exponential distributions. 
\item 
For sufficiently large $R_m$, the current sheets fragment along their length, leading to a sharp cutoff in the current sheet length distribution. The cutoff is consistent with the threshold for non-linear tearing/plasmoid instability.
\item 
For increasing $R_m$ the magnetic field lines become increasingly tangled, the mean and peak values of $|{\bf B}_\perp|$ increase, and the Poynting flux into the domain increases. This implies an increased heating rate.
\item 
The global reconnection rate is essentially independent of $R_m$, {though the energy dissipation rate increases with $R_m$}.
\item 
For a magnetic Prandtl number ($Pr_m$) of $\sim 1$ the Joule dissipation dominates the viscous dissipation by a factor of $\sim 3$. Note that this may depend on more than just $Pr_m$: in our simulations the ``free'' magnetic energy is around a factor of 2 higher than the kinetic energy, but this is different in other literature studies.
\end{enumerate}
All of the above suggest that the braiding mechanism can effectively heat the internal portions of coherent flux tubes in the corona. In the presence of the Sun's fragmented photospheric flux distribution, the coronal tectonics mechanism \citep{priest2002} will certainly also play a role in heating at the interfaces of such flux tubes, and indeed recent results by \cite{breu2025} show the two processes working in tandem. 
In the future it will be important to further explore the high-$R_m$ regime as here we have only just accessed the regime with tearing-unstable current layers. This is likely to require a modified computational approach at present levels of computing power. It is also necessary to explore how these results change when thermal conduction, radiative losses, and atmospheric stratification are included, as well as the clumpy nature of the photospheric flux distribution. These are left to future work.

\begin{acknowledgments}
D.P.~gratefully acknowledge support through Australian Research Council Discovery Projects (DP210100709, DP230101240), as well as helpful discussions with L.~P.~Chitta.
JAK was supported by the Heliophysics Internal Scientist Funding Model  (internally competed grant program) at Goddard Space Flight Center.
\end{acknowledgments}

%

\software{numpy \citep{numpy},  
          matplotlib \citep{matplotlib}, 
          opencv \citep{opencv_library}, scikit-image \citep{scikit-image}, UFiT \citep{aslanyan2024,ufit}
          }


\appendix

\section{Boundary driving}\label{app:driver}

The driving flow on the $z$-boundaries takes the form below: 
\begin{eqnarray}
    v_{x\mp} &=& v_0\frac{\sqrt{2L}}{\exp(-1/2)}y\exp\left( -L(x\mp x_0)^2-Ly^2 \right)\\
    v_{y\mp} &=&- 
    v_0\frac{\sqrt{2L}}{\exp(-1/2)}(x\mp x_0)\exp\left( -L(x\mp x_0)^2-Ly^2 \right)\\
    f_1(t) &=& \sum_j \frac{1}{2}(\tanh(a(t-b-2jT))-\tanh(a(t-T+b-2jT))\\
    f_2(t) &=& -\sum_j \frac{1}{2}(\tanh(a(t-b-T-2jT))-\tanh(a(t-2T+b-2jT))
\end{eqnarray}
and
\begin{eqnarray}
    v_x &=& f_1(t)v_{x-}+f_2(t)v_{x+},\qquad
    v_y = f_1(t)v_{y-}+f_2(t)v_{y+},\qquad \textrm{on }~z=-10\\
    v_x &=& f_2(t)v_{x-}+f_1(t)v_{x+},\qquad
    v_y = f_2(t)v_{y-}+f_1(t)v_{y+},\qquad\textrm{on }~ z=+10,
\end{eqnarray}
with $x_0=0.1875$, $L=8$, $v_0=0.0204$, $T=40$, $a=4$, $b=0.75$. 

\section{Current sheet detection and properties algorithm}\label{app:edge}

\begin{figure*}
    \centering    \includegraphics[width=0.98\linewidth]{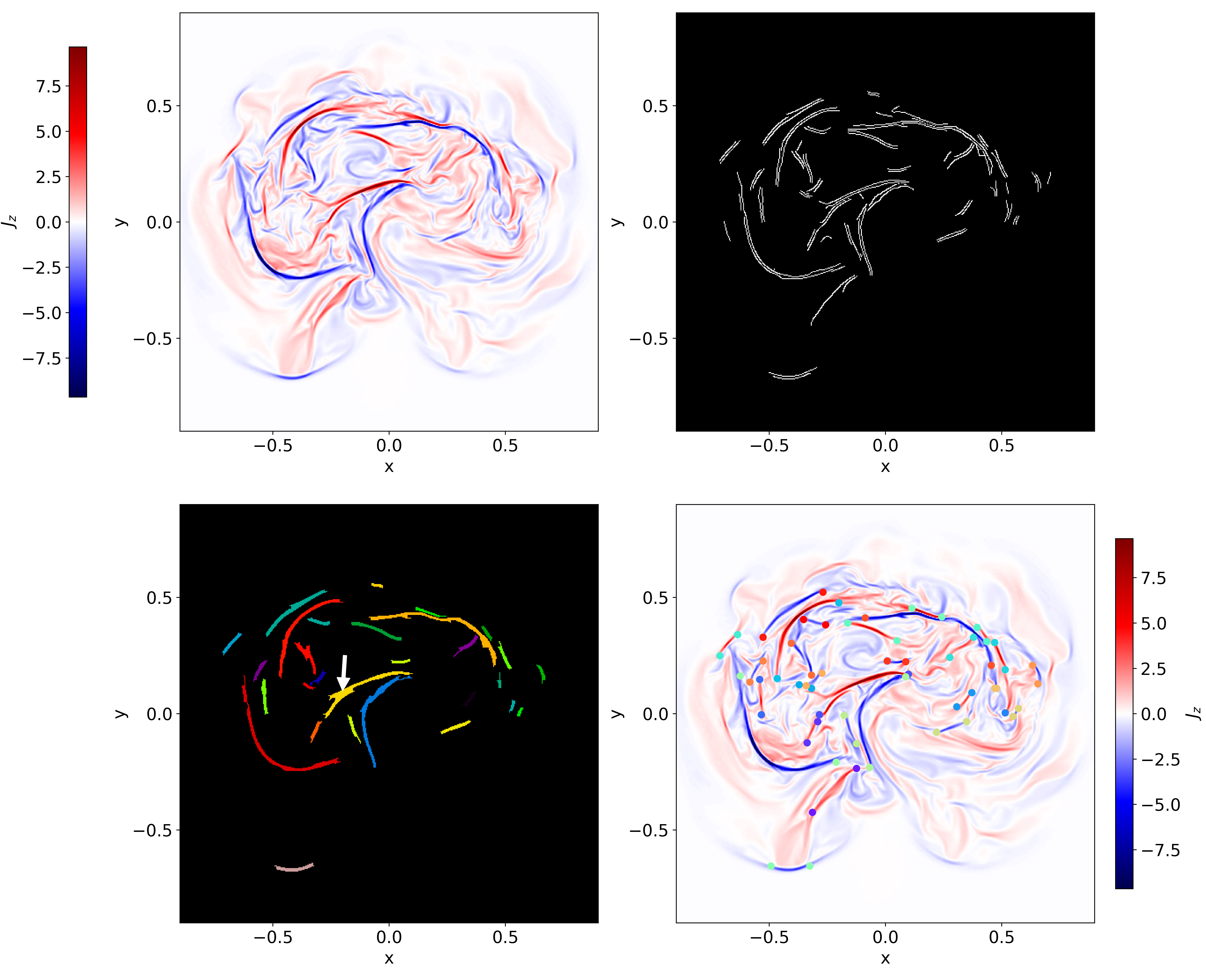}
    \caption{Illustration of the current sheet detection algorithm. Top-left: $J_z$ at $t=365$ for the medium resolution simulation. Top-right: edge pixels detected by the Canny edge detection algorithm with lower threshold of 100, upper threshold of 200, and with L2 norm option for the gradient evaluation.   Bottom-left: discrete current sheets identified -- different colours correspond to different current sheet labels. The white arrow marks a pair merged structure, described in the text. Bottom-right: $J_z$ distribution with the identified current sheet ends marked by pairs of coloured dots.}
    \label{fig:algorithm}
\end{figure*}

The first step of the current sheet detection algorithm involves finding `edges' in the current density distribution, for which we use Canny edge detection \citep{canny1986}, as implemented in the \texttt{opencv} library of \texttt{python}. In Figure \ref{fig:algorithm} we provide an example of the application of this and the following steps of the algorithm for a representative snapshot of the medium resolution simulation. Frame (a) shows the underlying current density, and frame (b) the detected edge pixels. Note that there are various options that can be tuned, for example to include more or fewer pixels in the edges. In order to balance the competing demands of (i) identifying most of the apparent current sheet edges versus (ii) not having the edges of adjacent current sheet merge, we have chosen fairly standard upper and lower threshold values of 200 and 100, respectively. Our results are found to be insensitive to the particular values chosen, so long as they are fairly close to these values which are typical values found in the literature. (Note that edge pixels are identified on the array $|J_z|$, rescaled to take values between 0 and 255.) The gradient is calculated using the L2 norm option. Following identification of edge pixels (Figure \ref{fig:algorithm}, top-right), the following steps were taken:
\begin{enumerate}
    \item 
    Separate the edge pixels into two groups with $J_z>0$ and $J_z<0$. Steps 2--5 are then performed for each of these sets of pixels individually.
    \item 
    With the goal to `grow' pairs of edges into sheets, perform a morphological closing operation. This is done using the \texttt{morphologyEx} function in \texttt{opencv}, with a $3\times 3$ square kernel. Note that in some cases, when same-sign current sheets come very close together, this may lead to two structures that appear by eye to be separate being identified as a single sheet. One such example is identified by the white arrow in Figure \ref{fig:algorithm}. We have optimised the size of the kernel to minimise such instances while also adequately joining the opposing edges of current sheets.
    \item 
    Remove small holes in connected groups of pixels -- holes with fewer than 64 pixels are filled.
    \item 
    Remove small objects -- connected groups of pixels less than 32 pixels in size are removed.
    \item 
    Label distinct, disconnected groups of pixels using \texttt{measure.label} from the \texttt{skimage} library.
    \item 
    Merge the two sets of labels. 
\end{enumerate}    
The resultant set of labelled regions for our sample snapshot is shown in the lower-left panel of Figure \ref{fig:algorithm}. From this overall list of labelled regions (i.e., current sheets), we perform the following steps to extract their properties:
\begin{enumerate}
\setcounter{enumi}{6}
    \item 
    Calculate the distances between all pairs of pixels, and find the two end points (maximally separated points) and the centre of mass (pixel with the minimum of the summed distances). The pairs of end points are shown in the lower-right panel of Figure \ref{fig:algorithm}.
    \item 
    Locate the pixel with the peak current.
    \item 
    Estimate the length of the current sheet as the summed distance from the first end point to the centre of mass to the peak current pixel to the second end point (note that there are two way to do this, and the smallest value of this distance should be chosen). 
    \item 
    Define two outflow speeds, $v_{\rm out}$, as the maximum values of $|{\bf v}_{xy}|$ within $10\times 10$ pixels of each end point.
    \item 
    Define the outflow direction at each end of the sheet using the end points and centre of mass. Find a unit vector perpendicular to this (in the $z=0$ plane) at each end and take the scalar product with ${\bf B}$ to get the outflow magnetic field component. Define $B_{\rm out}$ as the maximum values within $10\times 10$ pixels of each end point.
    \item 
    Define a single inflow field strength, $B_{\rm in}$, as the maximum value of $|{\bf B}_{xy}|$ within $20\times 20$ pixels of the peak current pixel. Note that the distribution of $|{\bf B}_{xy}|$ can be quite asymmetric across a current sheet -- see Figure \ref{fig:distributions} -- so that the distribution of $B_{\rm in}$ is missing some smaller values and the jump in $|{\bf B}_{xy}|$ will be less than $2B_{\rm in}$.
    \item 
    Approximate a vector tangent to the sheet at the midpoint using the two end points. Find a unit vector perpendicular to this and then take the scalar product with ${\bf v}$ to get the inflow velocity. Define $v_{\rm in}$ as the maximum value within $20\times 20$ pixels of the peak current pixel. Note that, as for $B_{\rm in}$, only the largest value from the two sides of the current sheet is identified, and these are often asymmetric.
\end{enumerate}

\section{Calculating the reconnection rate}\label{app:recrate}

\begin{figure*}
    \centering    \includegraphics[width=0.98\linewidth]{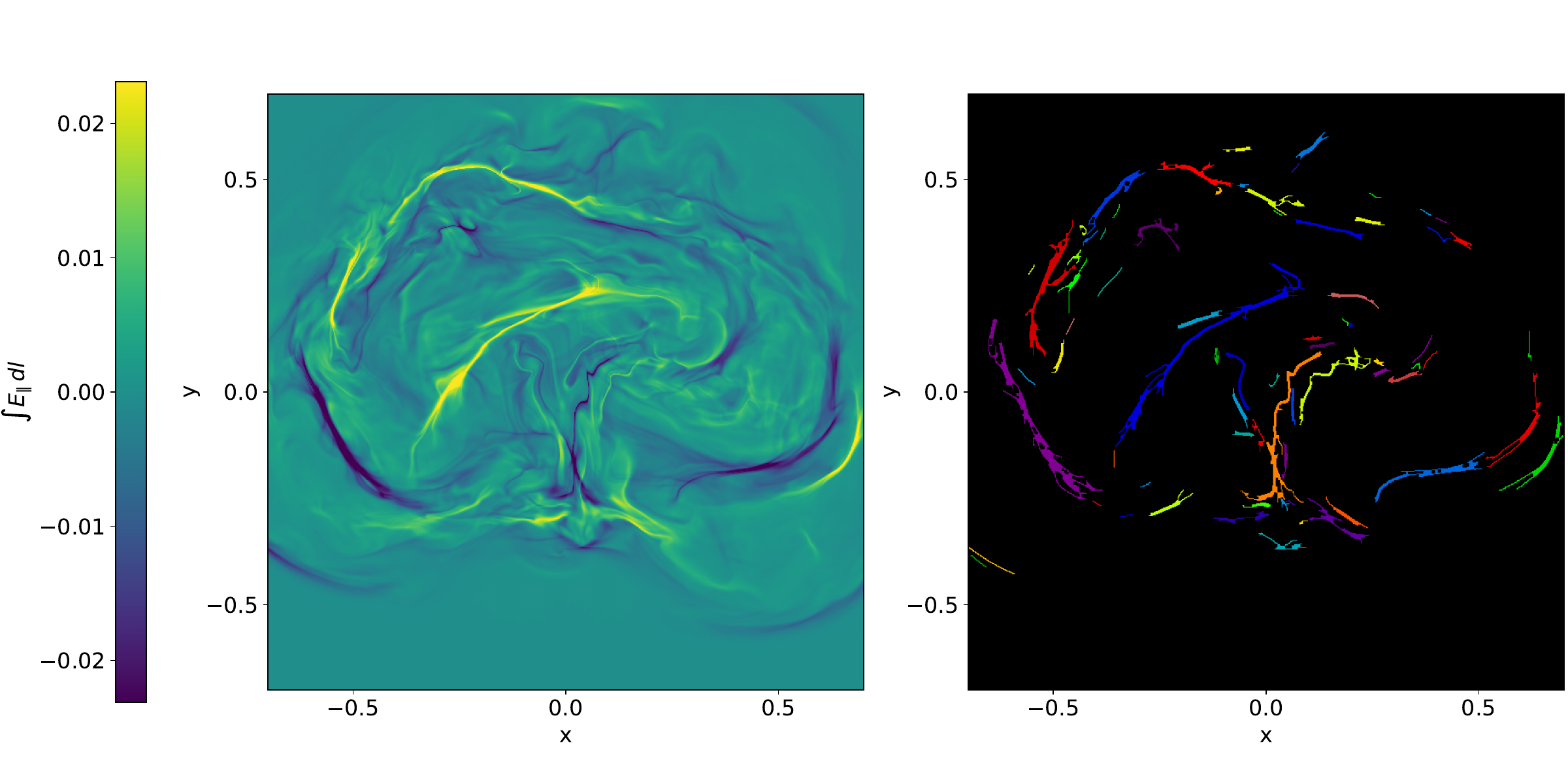}
    \caption{(a) $\Phi= \int E_\|\, dl$ plotted on the $z=0$ plane of the simulation with resolution $528^2\times 1280$ at $t=300$. (b) Labelled regions identified in the same quantity using the method and parameters described in the text.}
    \label{fig:int_epar}
\end{figure*}

To calculate the reconnection rate at a given time we start by calculating  $\Phi= \int E_\|\, dl$ along a set of field lines with seed points from a regular grid in the mid-plane $z=0$ ($800^2$ points equally spaced for $x,y\in[-0.8,0.8]$), where $l$ is the arc-length along the field line.  This is done by first calculating the dissipative electric field used internally within the code, using the formulation described in Equation (9) of \cite{gudiksen2011}. The step-size is set to ensure that each field line is resolved with at least 2000 points, and $E_\|$ is interpolated to these points using trilinear interpolation.

Having done this we wish to identify local maxima in $|\Phi|$. $E_\|$ is localised within current sheets, and so when we integrate along field lines, $\Phi$ also exhibits thin, sheet-like features as shown in Figure \ref{fig:int_epar}. Often these appear to be bifurcated as multiple sheets overlap when projected along the field lines \tb{(so that the corresponding field lines are reconnecting at more the one location along their length)}. There are many different ways to perform region identification and therefore identification of local maxima, but due to the sharp, elongated nature of concentrations of $\Phi$ we use the same process of identifying edges and then growing them into features as used for current sheet identification (see Appendix \ref{app:edge}). Due to the tendency of features to merge/overlap due to projection effects, we choose to use different values of various parameters to produce the plots in Figures \ref{fig:algorithm} and \ref{fig:int_epar}. These are chosen so as to be more likely to identify features as separated, compared to the parameters chosen for the current sheet identification. An example of the region segmentation is shown in Figure \ref{fig:int_epar}. For Figures \ref{fig:algorithm} and \ref{fig:int_epar} we used lower and upper thresholds for edge detection of 50 and 100, respectively, with thresholds for removing small objects and filling small holes of 16 pixels. It turns out that the value of the reconnection rate at any given time is sensitive to these choices, but not the qualitative trend over time presented in Figure \ref{fig:recrate}, or the fact that the values for different resolution ($R_m$) all lie close to one another.

{It is worth noting that there are a number of reasons why the numerical values obtained by the above procedure will be an under-estimate of the reconnection rate. These include (i) the overlapping of features mentioned above, (ii) the exclusion of the contribution of small regions (current sheets), (iii) the tendency of field lines to diverge away from the hyperbolic field regions at the heart of the current sheet, where $E_\|$ will typically be maximised. However, these issues would be expected to afflict the analysis of all three simulations, and it is the relative, not absolute, values of the reconnection rate that are of interest here.}


\bibliography{bibliog}{}
\bibliographystyle{aasjournalv7}



\end{document}